\documentclass[prl,aps,twocolumn,showpacs,floatfix]{revtex4}

\usepackage{epsfig}
\usepackage{amssymb}
\usepackage{amsmath}
\usepackage{amsfonts}
\usepackage{epstopdf}
\usepackage{verbatim}
\usepackage{color}
\usepackage[normalem]{ulem}
\begin{document}

\title{Simple mechanisms that impede the Berry phase identification from magneto-oscillations.}

\author{ A.\,Yu.~Kuntsevich$^{a,b}$\thanks {e-mail:
alexkun@lebedev.ru},
A.V. Shupletsov$^{a,c}$,
G.\,M.~Minkov$^{d,e}$}

\affiliation{$^a$P.\,N.~Lebedev Physics Institute, 119991 Moscow, Russia}
\affiliation{$^b$National Research University Higher School of Economics, Moscow 101000, Russia}
\affiliation{$^c$Moscow Institute of Physics and Technology, Moscow 141700, Russia}
\affiliation{$^d$Institute of Natural Sciences, Ural Federal University, 620000 Ekaterinburg, Russia}
\affiliation{$^e$M.~N. Miheev Institute of Metal Physics of Ural Branch of
Russian Academy of Sciences, 620137 Ekaterinburg, Russia
}

\begin{abstract}
Phase of quantum magnetooscilalions is often associated with the Berry phase and is widely used to argue in favor of topological non-triviality of the system (Berry phase $2\pi n+\pi$). Nevertheless, the experimentally determined value may deviate from $2\pi n+\pi$ arbitrarily, therefore more care should be made analyzing the phase of magnetooscillations to distinguish trivial systems from non-trivial.  In this paper we suggest two simple mechanisms dramatically affecting the experimentally observed value of the phase in three-dimensional topological insulators: (i) magnetic field dependence of the chemical potential, and (ii) possible non-uniformity of the system. These mechanisms are not limited to topological insulators and can be extended to other topologically trivial and non-trivial systems.
\end{abstract}

\maketitle
\section{Introduction}
Emergence of topologically non-trivial systems, like graphene or three-dimensional (3D) topological insulators(TI) required an experimental method to indicate topological non-triviality, i.e. presence of the two-dimensional (2D) carriers with a Dirac spectrum. A simple proper characteristic is Berry phase, $\phi_B$, which is known in two limiting cases:  $2\pi n$ (topologically trivial) or $2\pi n+\pi$ (topologically non-trivial)\cite{ando}, where $n$ is an integer number. It was conjectured\cite{mikitik1999}, that if the system exhibits magnetooscillations (Shubnikov-de Haas or de Haas-van Alphen), the phase of these oscillations straightforwardly reflects the Berry phase\cite{xiao}.  This phase enters the energy of Landau levels (LLs) through the quasiclassical quantization condition:
\begin{equation}
S(\varepsilon_N,k)=(\frac{2\pi|e|B}{\hbar})(N+\frac{1}{2}-\beta)
\label{QCl}
\end{equation}
where $S(\varepsilon_N,k)$ is a cross-sectional area of the $N$-th LL orbit in $k$ space and the offset $\beta$ is equal to Berry phase divided by $2\pi$\cite{mikitik1999}. Quasiclassical equation is applicable for high numbers of LLs $N\gg 1$.

Since that it has become popular to determine this offset from the fan diagrams (Berry plots), with the x-axis being number of conductivity minimum and the y-axis being the corresponding inverted magnetic field. As a rule the points in this diagram follow the straight lines and cross the x-axis at certain point. In case of two-dimensional carriers (like 2D gas in quantum well or at the surface of 3D TI), if this point is integer the system is believed to be topologically trivial, if this point is half-integer - topologically-nontrivial. The magneto-oscillation phase considerations were widely applied to graphene and graphite\cite{novoselov,zhang}, 3D topological insulators\cite{taskinprl,analytis, taskin, xiong, Pan, ren,lang, xiong2013}, Rashba semiconductors\cite{Murakawa}, high temperature cuprate superconductors\cite{cuprates} and pnictides\cite{CaFeAsF}, Weyl\cite{BaMnSb2, NbAs} and Dirac\cite{cdas1} semimetals, black phosphorous\cite{blackP} and gray arsenic\cite{grayas}, transition metal dichalcogenides\cite{mote2} etc.

However, quite often even in well understood systems, like 3D TIs, the experimentally observed value of offset deviated from the expected 0.5 value\cite{analytis, taskin, taskinprl, xiong, Pan}. In order to explain these discrepancies, more elaborate theoretical analysis\cite{mikitik, taskin, wright, ozerin} suggested several mechanisms: Zeeman splitting, absence of the electron-hole symmetry, trigonal wrapping of the Fermi surface, etc. It turned out that Zeeman splitting (large effective $g-$factor) is the most realistic option to explain experimental data in bulk crystals of bismuth chalcogenides. However it remains unclear why this $g-$factor might be so much spread from sample to sample (2 to 70). This uncertainty motivated us to look for alternative explanation.

In our paper we do not modify the model Hamiltonian, rather we consider simple macroscopic mechanisms. In particular,  we discuss the effect of chemical potential (whether it is constant or changes with magnetic field) on the phase of the quantum oscillations and show that it's role might be decisive. Moreover, we show that indirectly assumed sample homogeneity is also crucial for correct extraction of magnetooscillation phase. While illustrated in 3D TI's, our arguments are applicable for various multi-component systems (e.g. semimetals) and should clearly be taken into account.

The magnetooscillations in 3D TIs are believed to be due to topological surface states (TSS) that are 2-dimensional. Let's first consider the phase of magnetooscillations in a single two-dimensional (2D) system like semiconductor quantum well or graphene.

\section{Qualitative pictures}
\subsection{Two-dimensional systems}
\begin{figure}
\vspace{0.1 in}
\centerline{\psfig{figure=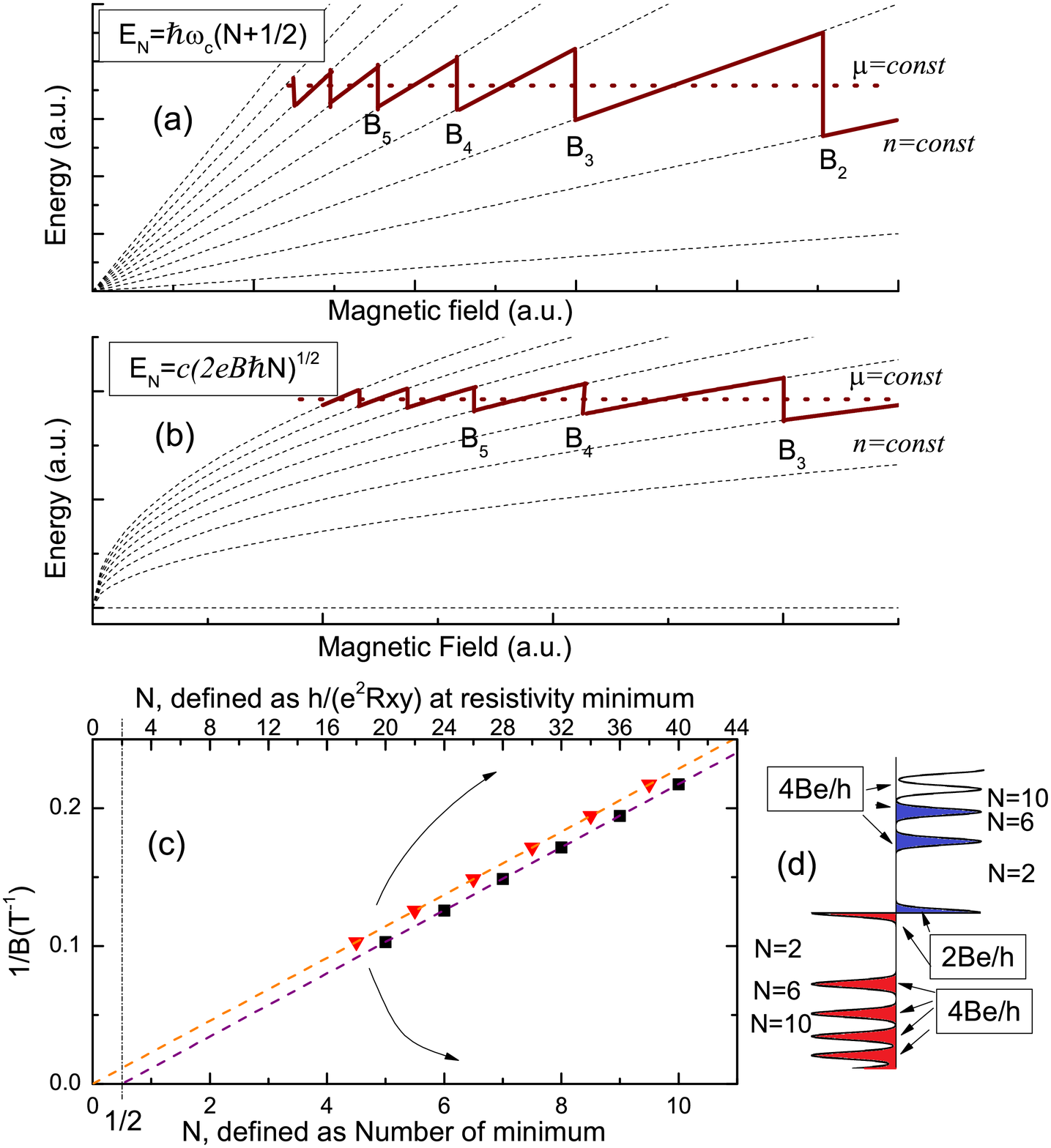,width=250pt}}
\begin{minipage}{3.2in}
\caption{(Color online) (a) Ladder of Landau levels versus magnetic field for spinless massive particles, solid brown line - chemical potential versus magnetic field provided that total density is fixed, according to Eq.\ref{2Deq}, dotted line - constant chemical potential, $B_N$ values are indicated; (b) the same as (a) for two-dimensional Dirac particles with linear dispersion; (c) fan diagram $1/B_N$ versus $N$ adopted from Ref.\cite{novoselov}. Top and bottom axes indicate two ways to define $N$(see text); (d)Landau level ladder in monolayer graphene for explanation of the anomalous phase of the magneto-oscillations.}
\label{fig1}
\end{minipage}
\vspace{0.4 in}
\end{figure}

It's a textbook knowledge that spectrum of two-dimensional system in perpendicular magnetic field $B$ consists of Landau levels (LLs) with a fixed degeneracy per spin $Be/h$ per unit area ($2.41\cdot 10^{10} \cdot B[{\rm T}]$ cm$^{-2}$). The overall electro-neutrality condition is reduced to constant total 2D electron density $n(T,B)=const$. Correspondingly, when integer number $N$ of LLs is filled, we get:
\begin{equation}
1/B_N=Ne/(hn)
\label{2Deq}
\end{equation}
This equation reflects the degeneracy of the LLs, and does not depend on zero-field spectrum of the carriers. Chemical potential traces the LLs and jumps across the gaps, at the points $B_N$, where Eq. (\ref{2Deq}) is fulfilled (See Fig. \ref{fig1}a). At these point minima in conductivity and resistivity are observed simultaneously and the integer number $N$ can be straightforwardly found from the Hall resistivity at the center of the $N$-th plateau $R_{xy}=h/(e^2N)$ in Quantum Hall Effect (QHE) regime. If one tries to determine the offset from the fan diagram of the 2D system, according to Eq.(\ref{2Deq}), one always has to get zero!!!
In case of Shubnikov-de Haas oscillations, i.e. if magnetic field is not high enough to open the complete gap between LLs, all these reasonings remain valid and $B_N$ values correspond to conductivity minima. A natural question arises: how is the nonzero Berry phase observed in graphene since the pioneering works (Refs.\cite{novoselov, zhang})?

In graphene the Dirac spectrum leads to square root dependency of the LL positions from the number $N$ and magnetic field $B$:

\begin{equation}
E_N=\pm\sqrt{2N\hbar eBv^2}
\label{3DTILL}
\end{equation}

Here $v$ is the speed of electrons assuming linear dispersion. This dependency is shown schematically in Fig.\ref{fig1}b. Each LL, including zeroth, has 4-fold degeneracy (2 spins $\times$ 2 isospins). 
Zeroth LL is half-populated by electrons (2-fold degeneracy) and half-populated by holes (2-fold degeneracy) (it is illustrated in Fig.\ref{fig1}d). In order to get non-trivial phase from magnetooscillations, one has to forget about the degeneracy of the levels and just count the minima of the resistivity. For example, the fan diagram, adopted from Ref.~\cite{novoselov} clearly shows the offset 0.5 (bottom axis, black boxes in Fig.\ref{fig1}c). If we define filling factors from the Hall resistivity as $h/(e^2R_{xy})$ (top axis, red triangles in Fig. \ref{fig1}c), we get crossing of the x-axis at zero in complete agreement with Eq.(\ref{2Deq}).

To sum up, the fan plot in graphene shows that zeroth LL has two times smaller degeneracy for electrons and this is a signature of the Dirac cone, that is related to non-trivial Berry phase. Moreover, there are only two possibilities for the values of the offset in case $n=const$ in any 2D system: integer (if there's no zeroth LL  equally shared between electrons and holes) and half-integer (if there is one). Actually, the latter is observed only in graphene, if the conductivity minima are counted, and only because the $N>1$ LLs are not further splitted by Zeeman effect.

\subsection{Three-dimensional topological insulators}
Three-dimensional topological insulators are much more common objects for speculations about the Berry phase. Zeroth Landau level for Dirac surface states of the 3D TI has $eB/2h$ degeneracy per unit of surface area, and contribution of the zeroth level should cause half-integer quantum Hall effect per one surface. Apart from graphene, in 3D TIs (i) the spectrum of the TSS has stronger deviations from the ideal Dirac one and (ii) there is large number of bulk states besides the surface carriers.

Many efforts were made to move the Fermi level of the 3D TIs into the gap and decrease the contribution of bulk carriers\cite{kushwaha}. However, at least in bismuth chalcogenides (the most studied 3D TIs), there are a lot of low mobility bulk states anyway  ($\sim 10^{17}$ cm$^{-3}$, see Refs. \cite{xiong,kushwaha,ren}; the density of bulk carriers is obtained from saturation of Hall effect in high field according to two band model) that do not experience Landau quantization in magnetic field, apart from surface states. They lead to large density of states on the Fermi level and therefore to pinning of the chemical potential.
Thus, in 3D TIs for TSS the $\mu=const$ condition seems to be realized instead of $n=const$ condition in 2D systems.

Let us illustrate what phase is expected in case $\mu=const$. In order to find conductivity {\it maxima} positions $B^{max}_N$ (when LLs cross the Fermi level) one has to solve Eq. (\ref{QCl}) with $\varepsilon_N=\mu$. For the Dirac spectrum this quasiclassical procedure leads to LLs, coincident with the exact solution, given  by Eq.( \ref{3DTILL}), and we get:

\begin{equation}
2N\hbar eB^{max}_N v^2=\mu^2=const
\label{DiracCond}
\end{equation}

It gives $1/B^{max}_N\propto N$. Conductivity minima are located at $B_N$ roughly in the middle between the corresponding maxima $B^{max}_N$ and $B^{max}_{N+1}$. Therefore, for the Dirac spectrum, the offset is equal to $0.5$. For the parabolic spectrum $\epsilon=p^2/2m$ the same procedure leads to the equidistant LLs $E_N = \hbar eB/m(N+1/2)$ and, correspondingly, zero offset of the fan diagram, because the conductivity minima correspond to the Fermi level between LLs:

\begin{equation}
\hbar \frac{eB^{min}_N}{m} N=\mu=const
\label{ParabolCond}
\end{equation}

To summarize the above qualitative considerations, $n=const$ condition makes the offset value sensitive only to LL degeneracy and insensitive to the spectrum of carriers. Indeed, Berry plot in this case is usually obtained from Eq. \ref{2Deq} which doesn't depend on positions of LLs (defined by spectrum of the carriers through Eq.\ref{QCl}). Since in 2D systems all LLs with  $N>0$  have the same degeneracy $eB/h$, the offset value for $n=const$ depends only on whether there is zeroth LL (graphene) with degeneracy $eB/2h$ or not (conventional 2D systems). The opposite $\mu=const$ condition, according to Eq.(\ref{QCl}), makes the offset sensitive to Berry phase and spectrum details. Besides, $\mu=const$ condition (unlike $n=const$) distinguishes low LLs (1,2,3) and high LLs (5,6,7...) as Eq.\ref{QCl} applicable only for $N\gg 1$. In two limiting cases: massless Dirac system and spinless system with the parabolic dispersion, the textbook values (0.5 and 0, respectively) of the offset are reached. Interestingly, these values exactly coincide with $n=const$ case (in Figs. \ref{fig1}a,b dotted lines $\mu=const$ cross sawtooth-like solid lines $n=const$ approximately in the Landau midgaps).
However, in general, there is no coincidence, and in $\mu=const$ case the deviations of the spectrum from ideal Dirac one lead to deviations of the Berry plot offset from 0.5 (for 3D TIs see e.g. Refs.\cite{taskin,wright,ozerin,mikitik}).

 Considered conditions are realized only in ideal 3D TIs. Indeed, $n=const$ is applicable only for pure 2D system without any excess reservoir of carriers, i.e. in 3D TIs it can be realized only if chemical potential lies in bandgap where there are no bulk carriers at all, that can't be reached in real 3D TIs. $\mu=const$ condition is realized only in 3D TIs with continuous density of bulk states on the Fermi energy level which is much larger than density of surface states. However, real 3D TIs may deviate from $\mu=const$ limiting case.

\section{Deviations from $\mu=const$ case}
\subsection{Crossover from $n=const$ to $\mu=const$ in 3D TI thin films}
We first consider thin films, as intermediate case, where the crossover from $n=const$ to $\mu=const$ might be realized. Indeed, the density of the impurity band states for bulk carriers is small due to negligible thickness. Correspondingly, when the total electron density is varied by gate voltage, the Fermi Level is tuned from gap (where only TSS are present and $n=const$) to valence (VB) or conduction band (CB) (where bulk states provide $\mu=const$ condition). Apparently, one would expect differences between these two limits only if the Hamiltonian deviates from the ordinary Dirac one. In the most popular family of 3D TIs Bi$_{2-x}$Sb$_{x}$Se$_{3-y}$Te$_y$ mainly Zeeman term (introduced in Ref.\cite{Hamiltonian}) was shown to affect the phase significantly\cite{taskin}:
\begin{equation}
\hat{H}= v(k_y\hat{\sigma}_x-k_x\hat{\sigma}_y) + 0.5g\mu_B B_{z}\hat{\sigma}_z
\label{zeemanterm}
\end{equation}

Here ${\bf k}=(k_x,k_y)$ is a quasimomentum vector, that should be replaced by ${\bf k}-e{\bf A}/c$ in magnetic field, $g$ is effective g-factor, $\mu_B$ is Bohr magnetron, ${\bf \sigma}$ is a vector of Pauli matrices. The spectrum of LLs is modified in the following way \cite{taskin}:
\begin{equation}
E_N=\pm\sqrt{2N\hbar eBv^2 + \left(\frac{g\mu_B B}{2}\right)^2}
\label{3DTILLZeeman}
\end{equation}

Zeeman term deflects chiral electron spin structure out of the surface plane and becomes significant in large magnetic fields, leading to non-linearity of $1/B_N(N)$ dependence. For further estimates we take $v\sim 3\times 10^5$ m/s (for Bi$_2$Se$_3$) from the ARPES measurements\cite{xia}. In various previous papers \cite{analytis,taskin,Pan} an arbitrary offset was explained by tuning the value of $g$-factor. In our estimates we suppose a fixed moderate value $g=30$ and Gaussian LL broadening with $\Gamma=1$ meV, close to the theoretical predictions \cite{Hamiltonian}.

We suggest the following realistic toy model of the Bi$_2$Se$_3$ 3D TI thin film: the system consists of two surfaces, hosting Dirac fermions with the spectrum of LLs given by Eq.(\ref{3DTILLZeeman}) and 3D bulk states with field-independent density of states per unit area:
\begin{equation}
D_{3D}(E)=\frac{2m_{||}\sqrt{2m_\perp (E-E_0))}}{2\pi^2\hbar^3}d
\end{equation}
where $d$-is the film thickness, taken to be 10 nm, $m_{||}$ and $m_\perp$ are effective masses in-plane and perpendicular to plane, respectively, $E_0$ is the bottom of the conduction band position, calculated relative to Dirac point of the TSS, and taken to be equal 150 meV (half of band gap in Bi$_2$Se$_3$). The effective masses are taken from ARPES/magnetotransport measurements\cite{lahoud} ($m_\perp \sim 0.25 m_e$, $m_{||} \sim 0.5 m_e$).
We also assume for simplicity that top and bottom surfaces of the 3D TI are equivalent.

The chemical potential and electron density must satisfy the following equations:
\begin{equation}
n_{3D}(B)+n_{2D}(B)=n=const
\label{electroneutrcond}
\end{equation}
\begin{equation}
\mu_{3D}(B)=\mu_{2D}(B)
\label{eqcond}
\end{equation}
The first one is the conservation of total charge ($n_{3D}$ and $n_{2D}$ are total densities of the 3D and 2D carriers in the film per unit area) and the second one is the thermodynamical equilibrium condition. In zero-temperature limit the corresponding densities are calculated as $n=\int^{\mu}_{0} D(\epsilon,B)d\epsilon$, where $D(\epsilon,B)$ is the density of states per unit area.

\begin{figure}
\centerline{\psfig{figure=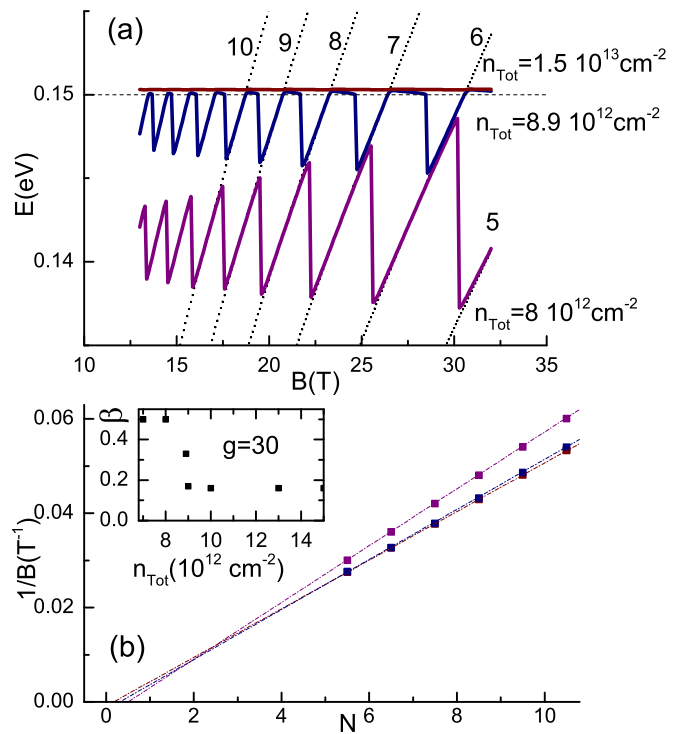,width=250pt}}
\begin{minipage}{3.2in}
\caption{(Color online) (a) Landau levels (5th-10th) versus magnetic field for TSS in model 3D TI Bi$_2$Se$_3$ thin film with effective $g$-factor equal to 30. Chemical potential versus magnetic field provided that total density is fixed,for three values of the total density; (b) the same as (a) for two-dimensional Dirac particles with linear dispersion; (b) the corresponding fan diagrams $1/B_N$ versus $N$ . The inset show schematically the dependence of the phase factor on total carrier density.}
\label{fig4}
\end{minipage}
\vspace{0.4 in}
\end{figure}

Solving them, we get the dependence of Fermi Level on magnetic field, find the intersection of $\mu(B)$ (Fig. \ref{fig4}a) and N-th Landau midgap ($B_N$) and build the dependence $N(1/B_N)$ (Fig. \ref{fig4}b). In order to avoid dimensional parameters (like value of the magnetic field in T or carrier density in units of $10^{12}$cm$^{-2}$) we use the oscillations numbers from 5 to 10.

 For low total density, when Fermi Level is in the gap, the offset value is equal to 0.5, as expected. For high density, when Fermi level is deep in the CB, intercept is -0.5 but in crossover regime intercept can be even less! Thus, if the Zeeman splitting is strong enough, an arbitrary phase can be achieved in the crossover between two limiting cases $n=const$ and $\mu=const$.

 An experimental realization of the suggested mechanism would be total density dependent  (i.e. gate-voltage dependent) offset value.  We should note, however, that there are almost no reported magnetooscillations data in thin films of (Bi,Sb) chalcogenides, where Fermi level is tuned across the gap.

We believe our considerations are supported experimentally by Ref. \cite{lang}, where the offset value changes from 0.5 to 0.2 as the chemical potential level is moved from the band gap to conduction band (see Fig. 4b of the above reference).

\subsection{Linear dependence of $\mu$ with field}
\begin{figure}
\vspace{0.1 in}
\centerline{\psfig{figure=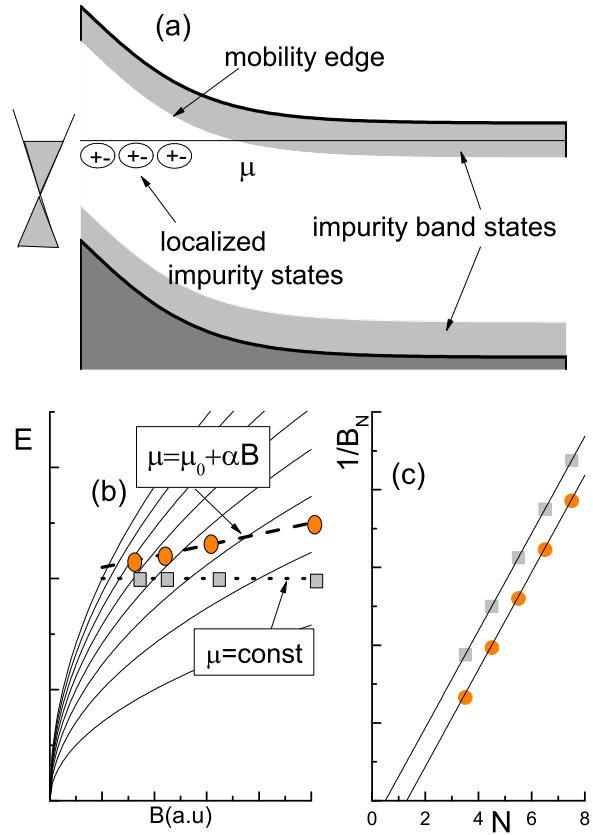,width=230pt}}
\begin{minipage}{3.2in}
\caption{(color online)(a) Schematic band diagram of the low density 3D TI, adopted from Ref.\cite{xiong2013}. (b) Energy diagrams ($E_n(B)$ dependencies) for the topological surface states Landau Levels (solid lines). Dotted and dashed lines - chemical potential within two models. Gray bars and orange circles - correspond to Landau gaps.(c) Fan diagrams, corresponding to $\mu=const$ (gray bars) and $\mu=\mu_0+\alpha B$(orange circles)}
\label{munotconst}
\end{minipage}
\end{figure}

Another chemical potential-related mechanism of the effective offset shift can be realized in clean bulk 3D TI samples (see e.g. experiments \cite{xiong,ren, analytis}). Bulk states have much larger density of states and much smaller mobility than the TSS. Assume the chemical potential level drifts with magnetic field, e.g. $\mu(B) = E_0+\alpha B$ instead of $\mu(B) = const$, as shown in Fig.\ref{munotconst}b. In this case positions of the Landau gaps (circles in Fig.\ref{munotconst}b) become shifted with respect to $\mu(B) = const$ case (bars in Fig.\ref{munotconst}b). The corresponding offset also shifts, as shown in Fig.\ref{munotconst}c.

The effect of the chemical potential drift with respect to the Dirac point (zeroth LL) is demonstrated  experimentally in Fig. 2c of Ref.~\cite{yoshimiTunnel} from tunnel spectroscopy measurements (0th LL shifts with respect to chemical potential level as field increases). The drift velocity $d\mu/dB$ was about 1 meV/T. Let's estimate the offset shift, caused by such drift of the chemical potential. We put $\mu(B) = E_0+d\mu/dB B_N$ into Eq.\ref{QCl} instead of $\varepsilon_N$, and neglect the second-order in $d\mu/dB$ terms. For the parameters of $Bi_2Se_3$  ($E_0=150$ meV, $v=3\times 10^5$ m/s ), the offset has a shift $h^{-1}e^{-1}v^{-2}E_0 d\mu/dB\approx0.37$. This estimate naturally explains almost arbitrary value of the offset in magnetooscillation experiments in 3D TIs.

Where does this magnetic field dependence of the chemical potential comes from? We can suggest some scenarios. For example, if the disorder is weak enough, and Fermi level is pinned by the bottom of the conduction band (see Fig.\ref{munotconst}a), locally the spectrum of bulk states in the band tail might also be quantized in magnetic field. Assume that 3D density can be so low, that starting from certain magnetic field all bulk electrons are placed in zeroth bulk LL, and acquire minimal additional energy $\hbar\omega_c/2$.
For the realistic mass of the 3D carriers ($m_\perp\approx0.25m_e$ in Bi$_2$Se$_3$), we get a reasonable value $d\mu/dB\sim 0.3$ meV/T. Another, even stronger effect might be the Zeeman drift of the chemical potential in the band tail. Indeed, if spin-orbit-interaction-renormalized $g-$factor is large enough, then the chemical potential of the band carriers should decrease with field. Yet another scenario, also believable in such narrow-band semiconductors as bismuth chalcogenides, is the sensitivity of the band gap and overall spectrum to magnetic field, due to magnetic field effect on atomic levels, Bloch functions, etc. We believe, eventually the nature of the non-zero $d\mu/dB$ value will be clarified.

\section{Influence of inhomogeneities}
Another issue, that can cause incorrect treatment of the magnetooscillations data is the choice of criterion of the integer number of the filled LLs. Which feature should be associated with Landau gaps: maximum or minimum of the resistivity (or anything else)?
The generally accepted answer was given by Xiong et al \cite{xiong} that conductance minima coincide with Landau gaps because conductance is defined by conductivity of the system.
Noteworthy, conductance (global characteristic) corresponds to conductivity (local characteristic) only in homogeneous system. Let's illustrate how the inhomogeneity can make the oscillation phase misleading.

\begin{figure}
\vspace{0.1 in}
\centerline{\psfig{figure=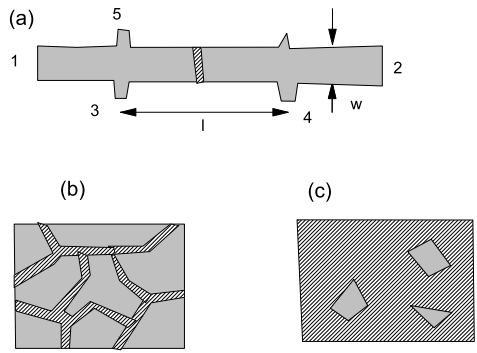,width=230pt}}
\begin{minipage}{3.2in}
\caption{(a)The simplest case of nonuniform sample (long rectangular sample in Hall bar geometry with crack, shown by hatching). Current flows between contacts 1 and 2. Longitudinal resistivity is measured between contacts 3 and 4 and contain contribution of cracks. Hall resistance is measured between contacts 3 and 5 and corresponds to bare material; (b),(c) other possibilities of non-uniformity.}
\label{nonuniform}
\end{minipage}
\end{figure}

Consider the simplest stripe-shape system constructed by homogeneous single-component high mobility parts separated by highly resistive transition region, e.g. one crack or grain boundary(Fig. \ref{nonuniform}a).

For uniform regions the minima of conductivity correspond to minima of local resistivity, because oscillations are always observed for classically large magnetic fields $\mu B>1$

\begin{equation}
\sigma_{xx}=\frac{ne\mu}{1+\mu^2B^2}\propto\frac{\rho_{xx}}{\rho_{xy}^2}
\end{equation}
Thus Landau gaps correspond to minima of local resistivity $\rho_{xx}$.
Total 4-wire resistance (inverse conductance) equals to sum $\rho_{xx}l/w+R_{C}$, where $R_{C}$ is the resistance of the crack. At the same time Hall effect is unaffected by the crack. If $R_{C}$ is so large that exceeds the Hall resistance in relevant magnetic fields, than the effective conductance $G$ can be evaluated as:

\begin{equation}
G_{xx}\approx \frac{1}{\rho_{xx}+R_Cw/l}
\end{equation}

Thus, conductance minima correspond to {\it maxima} of resistance (and therefore to maxima of conductivity) and phase of oscillations acquires artificial $\sim\pi$ shift. Depending on the configuration of non-oscillating regions (Fig. \ref{nonuniform}b,c), an arbitrary shift can be received. How can the presence of such regions be indicated in real samples? The most reliable and expensive approach would be detailed microanalysis. There is, however, an indirect indicator.

It is a textbook knowledge \cite{dassarma} that mobility derived from Shubnikov-de Haas oscillations ($\mu_{SdH}\equiv e\tau_D/m\*\approx1/B_{ons}$, where $\tau_D$ and $m^*$ are Dingle time and effective mass correspondingly, and $B_{ons}$ - is the minimal magnetic field where oscillations emerge) can't exceed the one from Hall coefficient ($\mu_{Hall}\equiv {\rho_{xx}}^{-1}d\rho_{xy}/dB$), especially in Dirac systems where backscattering is prohibited. Experimentally, however, the opposite relation is often observed $\mu_{SdH}>\mu_{Hall}$ in 3D topological insulator systems\cite{ren,xiong, taskinprl,xiong2013}.
This anomalous ratio was attributed to huge reservoir of low-mobility bulk carriers.
Interstingly, even in the first research, where the conductance criterion was suggested \cite{xiong}, the offset values determined from $G_{xx}$ and $G_{xy}$ disagree with each other, thus showing up incompleteness of the multi-liquid model.
In thin films of 3D TIs ($\sim 10-40$ nm ) it is hard to imagine a huge reservoir of low mobility carriers, while the $\mu_{SdH}/\mu_{Hall}$ ratio may exceed one\cite{jauregui,tu}.

On the contrary, such ratio can be explained in all systems by presence of transition regions. Indeed, if transition regions are responsible for this high resistivity, whereas low-disorder domains provide intensive magnetooscillations starting from relatively low fields, this high $\mu_{SdH}/\mu_{Hall}$ ratio is naturally explained. However, if the sample is inhomogeneous, it is becoming absolutely unclear which criterion  for the integer LL index  should be used.

For example, in Ref. \cite{tu} physical vapor deposition grown Bi$_{2-x}$Sb$_x$Te$_{3-y}$Se$_y$ 18-nm thickness films are reported with Hall mobility less than 30 cm$^2/$Vs and Shubnikov-de Haas mobility in the range between 2500 and 5000 cm$^2$/Vs. In the same films relative amplitude of the Shubnikov-de Haas oscillations was less than 0.01\% (0.2 Ohm atop of 2400 Ohm), thus clearly signifying the case, shown in Fig.~\ref{nonuniform}c.

\section{Discussion}
On the basis of the discussed three mechanisms we suggest that the best object for studies of the phase of quantum oscillations in 3D TIs would be thin films or flakes, because of negligible density of the in-gap states. If there is no abnormal $\mu_{SdH}/\mu_{Hall}>1$ ratio, than most probably the system is uniform, and conductivity minimum criterion should be trusted. Indeed, this signature of the uniformity is present in most experimental papers on Shubnikov de Haas oscillations in thin films or flakes, where $\sim0.5$ offset is reported.

Another source of mistake in determination of the offset is the procedure of the straight line $1/B_N(N)$ plotting. If the typical numbers $N$, used for fit, are large (about 10-25), and amount of minima is small (5-7), than the mistake can be essential. We believe therefore, that for reliable statements about the offset value, the results should be demonstrated not on a single sample, but rather on a series of the samples.

Chemical potential drift mechanism, suggested by us, poses a question whether Zeeman term (see Eq. \ref{zeemanterm}) used for explanation of the anomalous offsets in Refs.\cite{analytis,taskin,ren,xiong} is relevant in 3D TIs. Indeed, such term in the Hamiltonian emerges only in magnetic field, i.e. it is invisible e.g. by ARPES, and can not be confirmed from independent measurements. Instead, the deviations of the Berry plot offset from 0.5 are naturally explained within the chemical potential drift.

Concerning bismuth chalcogenides, we should also note, that in doped crystals (with carrier density $\sim 10^{20}$~cm$^{-3}$) with negligible TSS contribution, Shubnikov-de-Haas oscillations of bulk states are often studied. These oscillations are quasi-2D, because they originate from almost cylindrical Fermi-surface\cite{lahoud}. However they are often attributed to surface carriers\cite{golubkov,shrestha, jacs}. Moreover 0.5 offset is often detected and considered to be a fingerprint of Diracness. We should note that in almost all of these bulk crystals the ratio $\mu_{SdH}/\mu_{Hall}$ has an abnormal value, larger than one, and that the straightforward application of the Eq. (\ref{QCl}) requires additional knowledge about the spectrum of these systems.

\subsection {Phase of magnetooscillations in 3D TI strained HgTe.}

Recently, strained epitaxial layers of HgTe (from 50 to 100 nm) were shown to be 3D topological insulators\cite{hgte1,hgte2,hgte3}.  Apart from bismuth chalcogenides, this material has zinc blend structure, no Van-der Waals bonds and, hence, high structural quality advantaged from well-developed epitaxial technology. The main features of 3D TI HgTe (as compared to bismuth chalcogenides) are:
(i) much higher mobilities up to 10$^6$ cm$^2$/Vs and almost complete absence of the in-gap bound states;
(ii) very small band gap ($\sim 10$ mV);

The complete Berry plot in the 3D TI regime (see Fig.3d in Ref. \cite{hgte1}) is nonlinear: in low magnetic fields (high $N$) only one surface demonstrates oscillations due to elevated mobility and density. The role of the second surface is to stabilize the chemical potential. The phase of these oscillations is abnormal $\beta\approx 0.6$, in agreement with \cite{hgte3} . At high magnetic fields (low $N$), both top and bottom surfaces are quantized (case $n=const$), and $N$ follows Eq. (\ref{2Deq}), without any phase shift.

The Berry plots 2.2V-4.4V in Fig. 3c in Ref.\cite{hgte2} correspond 3D TI with inequivalent surfaces and demonstrate nontrivial phase of the quantum magnetooscillations. As Fermi level of the top surface moves to the valence band (2V figure), the phase of the oscillations shifts (similarly to our Fig.~\ref{fig4}) Thus, the arguments of our paper could be straightforwardly applied to HgTe systems.

We should note however, that thin film-based HgTe 3D TIs are not that simple. For example, the gap positions determined from capacitance and resistivity do not coincide. For adequate analysis of the Berry plots in these systems (top gate + top surface of 3DTI + bottom surface of 3DTI) one should additionally solve a Poisson equation. Indeed, in order to maintain chemical potential common as magnetic field is swept, a redistribution of carriers between top and bottom surfaces should occur, thus affecting the electrostatics of the whole system. These calculations are out of the scope of this paper and yet have to be done.

Interestingly, Ref.\cite{hgte2} probes density of mainly the top surface through the quantum capacitance. This method has advantages over resistive detection, it does not suffer from conductivity/resistivity criterion, possible sample inhomogeneities and detunes from parasitic bulk and second surface contributions. Application of the capacitive technique to the other 3D TI materials will help to understand whether conductivity/resistivity criteria determine Landau gaps.

\subsection{Phase of magnetooscillations in 3D multiband systems}

Apart from 2D systems and surface states of 3D topological insulators, discussed in this paper, 3D metals (or semimetals) are not gapped in magnetic field, because they preserve dispersion in a magnetic field direction. In particular, for quadratic spectrum one has:
\begin{equation}
E_N(k_z)= \frac{\hbar eB}{m_{eff}}(n+\gamma)+\frac{\hbar^2{k_z}^2}{2m_z},
\end{equation}
where $k_z$ is the wave vector of electron in the magnetic field direction, $m_{z}$, and $m_{eff}$ are effective masses in parallel and perpendicular to magnetic field directions, respectively. For systems with linear dispersion one has:
\begin{equation}
E_N(k_z)= \hbar c\sqrt{\frac{2Be}{2\pi\hbar}(n+\gamma+C^2sin ^2(\theta))+{k_z}^2}
\end{equation}

Here $c$ is the velocity of electrons, $C$ - is the material dependent parameter, equal to zero in Weyl metal and not equal to zero in Dirac metal, $\theta$ is the angle between magnetic field and a certain crystallographic direction.

Correspondingly, the density of states between LLs becomes non-zero, leading to the shift of the conductance minima out of the center of Landau gap. In order to calculate the magnetooscillations phase shift one usually considers only first harmonic of the oscillations\cite{shoenberg}, that is justified only for large $N$. There are other factors, that make magnetooscillation phase in 3D case less reliable. For example, a realistic modification of the spectrum in topological metals (introduction of electron-hole assymetry) causes significant shift of the phase\cite{weyl}.

The mechanisms, described in our paper may readily affect the phase in numerous multiband 3D materials, like cuprates, pnictides, topological semimetals (e.g. Dirac and Weyl), etc. Indeed, if at the Fermi level there are only few equivalent bands with coincident LLs , than $n=const$ condition should be applied. In the opposite limit, when besides electrons of interest there is large side density of states from the other subbands, $\mu=const$ condition becomes applicable. However,  $\mu\neq const$ case is also entirely possible, especially for thin film objects.

In fact, for any multisubband system a theoretical analysis, similar to ours, should precede the treatment of the Berry plot data: (i) spectrum of LLs should be calculated for each subband; (ii) chemical potential should be found for each magnetic field from equilibrium and electroneutrality conditions (Eqs. similar to (\ref{electroneutrcond}), and (\ref{eqcond})); (iii) positions of the corresponding Landau gaps should be found; and (iv) the corresponding criterion (minima of conductance or heat conductance or anything else) should be chosen and justified.

 Interestingly, recently very similar ideas were implemented to theoretical analysis of the phase of magnetooscillations in nodal line semimetals\cite{nodalline}, i.e. materials where instead of single Dirac point a nodal line is observed.

\section{Conclusion}
 To sum up, positions of the Landau gaps are determined by thermodynamics of the system and detected by resistivity. We demonstrate that besides topology such practical aspects, as possibility for sample inhomogeneity and thermodynamical constrains crucially affect the phase of magnetooscillations in 3D topological insulators. The situation when the phase is different from $\pi$ is entirely possible, even for Dirac-like carriers. Therefore, the phase of magnetooscillations, at least in most studied 3D TIs (bismuth chalcogenides) should not be generally used to prove the Diracness. Rather, magnetooscillations phase might be only complimentary to other measurements. Generalization of our ideas to other material systems, like Dirac and Weyl semimetals can also be performed.

The authors are very thankful to V.A. Volkov and D.A. Kozlov for discussions. The work is supported by Russian Science Foundation grant No 17-12-01544.

\end{document}